\documentclass[preprintnumbers,amssymb, nofootinbib, nobibnotes,aps,prd]{revtex4}
\usepackage{soul, color}
\usepackage{amsmath}

\usepackage{graphicx}
\usepackage{dcolumn}
\usepackage{bm}

\newcommand{\beq}{\begin{equation}}
\newcommand{\eeq}{\end{equation}}

\begin{document}

\title{Spinning particles, axion radiation, and the classical double copy}%

\author{Walter D. Goldberger}%
\author{Jingping Li}
\author{Siddharth G. Prabhu}
\affiliation{Physics Department, Yale University, New Haven, CT 06520, USA}
\date{\today}%

\begin{abstract}

We extend the perturbative double copy between radiating classical sources in gauge theory and gravity to the case of spinning particles.    We construct, to linear order in spins, perturbative radiating solutions to the classical Yang-Mills equations sourced by a set of interacting color charges with chromomagnetic dipole spin couplings.    Using a color-to-kinematics replacement rule proposed earlier by one of the authors, these solutions map onto radiation in a theory of interacting particles coupled to massless fields that include the graviton, a scalar (dilaton) $\phi$ and the Kalb-Ramond axion field $B_{\mu\nu}$.    Consistency of the double copy imposes constraints on the parameters of the theory on both the gauge and gravity sides of the correspondence.   In particular, the color charges carry a chromomagnetic interaction which, in $d=4$, corresponds to a gyromagnetic ratio equal to Dirac's value $g=2$.    The color-to-kinematics map implies that on the gravity side, the bulk theory of the fields $(\phi,g_{\mu\nu},B_{\mu\nu})$ has interactions which match those of $d$-dimensional `string gravity,' as is the case both in the BCJ double copy of pure gauge theory scattering amplitudes and the KLT relations between the tree-level $S$-matrix elements of open and closed string theory.

\end{abstract}

\maketitle

\section{Introduction}
\label{sec:intro}

Almost one decade ago Bern, Carrasco, and Johansson (BCJ) discovered remarkable relations between perturbative amplitudes in gauge and gravity theories~\cite{Bern:2008qj,Bern:2010ue,Bern:2010yg}.   The BCJ correspondence generates gravity amplitudes by applying a set of simple color-to-kinematics transformations to the $S$-matrix of gauge theory, once written in a suitable form.    This correspondence includes, as a special case, the $\alpha'\rightarrow 0$ limit of the earlier KLT relations~\cite{Kawai:1985xq} found in tree-level string theory, but generalizes them to much wider classes of field theories, both at tree and loop levels.   See~\cite{Carrasco:2015iwa} for a recent review of the literature.   

Given the relative simplicity of the gauge theory Feynman rules, the BCJ correspondence has made accessible the evaluation of high precision perturbative observables that would otherwise be intractable by direct calculation in gravity.   See~\cite{Bern:2017ucb} for a recent example at five loops, based on developments in~\cite{Bern:2017yxu}.   It is therefore natural to ask if a similar ``double copy'' structure also underlies, thereby simplifying,  the calculation of observables beyond the $S$-matrix.   This question was first analyzed in the work of refs.~\cite{Monteiro:2014cda,KS2,KS3} within the context of classical Kerr-Schild solutions to the Einstein equations, and further developed in refs.~\cite{KSothers}.    More recently, ref.~\cite{Goldberger:2016iau} showed that the classical double copy can be applied to the analysis of radiation from perturbative, time-dependent sources.   In particular, ref.~\cite{Goldberger:2016iau} showed that  the classical bremsstrahlung radiation fields in a certain theory of gravity can be obtained from a simpler gauge theory calculation by a set of color-to-kinematics replacement rules which are similar to those used in the case of amplitudes.   This result was later generalized~\cite{Goldberger:2017vcg} to radiation from a system of point sources in bound orbital configurations, analogous to the compact binary inspirals recently detected via gravitational radiation emission~\cite{TheLIGOScientific:2017qsa}.

Note that the classical gravitational radiation fields found in~\cite{Goldberger:2016iau,Goldberger:2017vcg} are not those of pure gravity.    Rather, they are those of a dilaton gravity theory consisting of a scalar (dilaton) $\phi$ and the graviton $h_{\mu\nu}$.    This is consistent with the BCJ double copy of pure gauge theory, which by degree of freedom counting,
\begin{equation}
A_\mu\otimes A_\nu = \phi\oplus h_{\mu\nu}\oplus B_{\mu\nu},
\end{equation}
is a theory that has the scalar field $\phi$ as well as the Kalb-Ramond~\cite{Kalb:1974yc} axion $B_{\mu\nu}=-B_{\nu\mu}$ in addition to the graviton.    In the calculation of~\cite{Goldberger:2016iau,Goldberger:2017vcg}, the role of the dilaton was to cancel the explicit dependence on the spacetime dimensionality $d$ from the pure gravity Feynman rules, as discussed in~\cite{Bern:1999ji} (see also a cryptic remark made earlier in~\cite{Scherk:1974mc}).   However, for the non-spinning point sources considered in~\cite{Goldberger:2016iau,Goldberger:2017vcg}, there is no classical radiation in the anti-symmetric channel.

The fact that the radiation in the mode $B_{\mu\nu}$ does not arise in the results of~\cite{Goldberger:2016iau,Goldberger:2017vcg} can be understood on the basis of symmetry.   In order to have radiation in the axion channel, the point sources must have linear couplings to $B_{\mu\nu}$.   However, in the absence of additional structure, it is impossible to write linear interactions with the particle worldlines that respect both diffeomorphism invariance as well as the gauge symmetry $\delta B_{\mu\nu} = \partial_\mu \zeta_\nu - \partial_\nu \zeta_\mu$ of the bulk action.    On the other hand, if the particles carry spin, a coupling to the field strength $H=dB$, of the form
\begin{equation}
\label{eq:ad}
\int dx^\mu S^{\nu\sigma} H_{\mu\nu\sigma},
\end{equation}
is allowed, and one would expect to find axion radiation in the double copy of gauge theory coupled to spinning point color charges.

In this paper, we extend the classical double copy to include particle sources with spin.    Starting in section~\ref{sec:glue} from a system of weakly coupled adjoint color charges $c^a$, with spin couplings $\int d\tau c_a S^{\mu\nu} F_{\mu\nu}^a$ to the gluon field strength, we compute radiation to linear order in the spins.    In section~\ref{sec:dc}, we apply the same color-to-kinematics replacements as in~\cite{Goldberger:2016iau} to obtain a gravitational double copy radiation field.   Unlike the spin-independent case, the double copy is only consistent for a specific value of the chromomagnetic coupling, corresponding (in $d=4$) to classical particles that carry a gyromagnetic ratio equal to Dirac's value $g=2$.    Only for this choice of parameters do we find a gravitational field that is consistent with Ward identities.    As we check explicitly in section~\ref{sec:dc}, this solution encodes axion radiation in a theory of particles with interaction as in Eq.~(\ref{eq:ad}), and a bulk Lagrangian which is of the form
\begin{equation}
\label{eq:st}
S_g = -2 m_{Pl}^{d-2} \int d^d x \sqrt{g} \left[R - (d-2) g^{\mu\nu} \partial_\mu \phi \partial_\nu \phi + {1\over 12} e^{-4\phi} H_{\mu\nu\sigma}^2\right],
\end{equation}
(see also~\cite{Luna:2016hge}) at least to the order in perturbation theory that we consider in this paper.  This is precisely the action for ``string gravity" at non-critical dimension $d$ (in the classical limit, where the ${\cal O}(\hbar)$ dilaton potential can be neglected).    It also matches the double copy of pure gluon amplitudes, which is suggestive of a relation at higher orders in perturbation theory between the classical color-to-kinematics rules proposed in~\cite{Goldberger:2016iau} and BCJ duality of the $S$-matrix.  

Here, we focus our attention only to the case of radiation in the axion mode.   The complete agreement between the double copy and Eq.~(\ref{eq:st})  in all radiation channels will be presented in a separate paper~\cite{JS}.   (In the case of pure gravity, the analogous bremsstrahlung process has been analyzed in refs.~\cite{JV}).   Taken together with the bound state results in~\cite{,Goldberger:2017vcg}, the spin corrections studied in this paper and in~\cite{JS} bring the classical double copy one step closer to making contact with astrophysically relevant~\cite{TheLIGOScientific:2017qsa} sources of gravitational radiation, although a systematic procedure for projecting out the unwanted dilaton and axion modes remains to be fully developed (in the case of purely spinless sources, progress in this direction was made in the recent paper~\cite{Luna:2017dtq}, which adapts techniques introduced in the context of scattering amplitudes in ref.~\cite{Johansson:2014zca} to the classical problem).  To keep our discussion self-contained, we provide a review of the classical spinning particle formalism that we use in this paper in appendix~\ref{sec:spapp}.


\section{Gluon radiation from spinning color charges}
\label{sec:glue}

We consider a system of classical spinning Yang-Mills color charges which interact and emit gluon radiation to infinity.   Each particle is described by a trajectory in spacetime $x^\mu(s)$, a spin angular momentum $S^{\mu\nu}(s)=-S^{\nu\mu}(s)$, and a color charge~\cite{sikivie} $c^a(s)$ transforming in the adjoint representation of the gauge group.    The interactions with the gauge field are encoded in an interaction worldline Lagrangian which is 
\begin{equation}
\label{eq:sint}
S_{int} = - g_s\int dx^\mu  c_a(\tau) A^a_\mu +{g_s\kappa\over 2} \int d\tau  c_a(\tau)S^{\mu\nu}(\tau) F^a_{\mu\nu}+\cdots,
\end{equation}
where $g_s$ is the gauge coupling, and the coefficient $\kappa$ determines the strength of the particle's chromomagnetic interaction.   We denote by $\tau$ the reparametrization invariant time coordinate along the particle worldline.   Note that the form of the interaction is valid for either massive or massless particles.   In the massive case, $\tau$ is proportional to the proper time along the worldline, but more generally it is related to an arbitrary worldline parameter by $d\tau(s) = e(s) ds$, where $e(s)$ is a  non-dynamical `einbein' inserted to ensure reparametrization invariance $s\rightarrow s'(s)$.   Only terms linear in spin, and with up to one derivative of the gauge field are kept in our analysis.    We have omitted kinetic terms for the degrees of freedom $x^\mu(s)$, $S^{\mu\nu}(s)$, $c_a(s)$, which are spelled out in more detail in appendix ~\ref{sec:spapp}.

The equations of motion for this system consist of the Yang-Mills equations\footnote{The conventions are $D_\mu = \partial_\mu + i g_s A^a_\mu T^a$,  $[T^a,T^b]=if^{abc} T^c$, $(T_{\mbox{\tiny{adj}}}^a)^b_c=-if_{abc}$.}
\begin{equation}
\label{eq:YM}
D_\nu F^{\nu\mu}_a(x) = g_s J_a^\mu(x),
\end{equation}
where the color current sourced by the point charges (labeled by the index $\alpha=1,2,\cdots$) is
\begin{equation}
\label{eq:ppcurrent}
J_a^\mu(x) = -{1\over g_s} {\delta\over \delta A^a_\mu(x)} S_{int} = \sum_\alpha \int dx_\alpha^\mu \,c^a_\alpha(\tau_\alpha)  {\delta(x-x_\alpha(\tau_\alpha))} -  {\kappa}_\alpha \int d\tau_\alpha \,S_\alpha^{\mu\sigma}(\tau_\alpha) D_{\sigma}\left[c^a_\alpha (\tau_\alpha) \delta(x-x_\alpha(\tau_\alpha))\right].
\end{equation}
The time evolution of the color charges then follows from the covariant conservation of this current $D_\mu J^\mu_a(x)=0$, which yields
\begin{equation}
\label{eq:ceom}
(v\cdot D) c^a = {ig_s\kappa \over 2}  [S^{\mu\nu} F_{\mu\nu},c]^a,
\end{equation} 
where we define ${v}^\mu = dx^\mu/d\tau$.

Likewise, the orbital equations of motion follow from the conservation of total energy-momentum, $\partial_\mu T^{\mu\nu} =0$, where $T^{\mu\nu}$ receives contributions from the gauge field and from the point particles themselves.    As reviewed in appendix ~\ref{sec:spapp}, it is necessary to impose a constraint on the spin $S^{\mu\nu}$ in order to reduce to the correct number of physical spin degrees of freedom implied by Poincare invariance.   We find it convenient to implement the choice
\begin{equation}
p_\mu S^{\mu\nu}=0
\end{equation}
which is sometimes referred to as the `covariant spin supplementary condition'.    With this choice, the energy-momentum tensor for a single spinning particle, defined by
\begin{equation}
T_{pp}^{\mu\nu}(x) = \int dx^{(\mu} p^{\nu)} {\delta(x-x(\tau))} + \int  dx^{(\mu}S^{\nu)\sigma}  \partial_\sigma {\delta(x-x(\tau))} -{\kappa g_s} \int d\tau {\delta(x-x(\tau))\over \sqrt{g}} c_a {F^a}_\sigma{}^{(\mu}{} S^{\nu)\sigma},
\end{equation}
is such that, for $F^{\mu\nu}=0$, the global momentum and angular momentum of the particle are 
\begin{eqnarray}
p^\mu &=& \int d^3 {\bf x} \, T^{0\mu}({\bf x},x^0),\\
J^{\mu\nu} &=& \int d^3 {\bf x} \, x^{[\mu} T^{0\nu]}({\bf x},x^0)= x^\mu p^\nu - x^\nu p^\mu + S^{\mu\nu},
\end{eqnarray}
as measured by a fixed inertial observer.   Given the form of the energy-momentum tensor, the equations of motion follow:
\begin{eqnarray}
{d\over d\tau} p^\mu &=&   g_s c^a F_a^{\mu\nu} v_\nu -{1\over 2}\kappa g_s  c_a S^{\lambda \sigma} D^\mu F^a_{\lambda \sigma},\\
{d\over d\tau} S^{\mu\nu} &=&  p^\mu {v}^\nu - p^\nu {v}^\mu - 2\kappa g_s  c^a F^{\lambda[\mu}_a S_\lambda{}^{\nu]}.
\end{eqnarray}
They imply in particular that $S_{\mu\nu} S^{\mu\nu}$ and  $m^2 = p_\mu p^\mu +  g_s\kappa c_a S^{\mu\nu}F^a_{\mu\nu}$ are conserved along the worldline.    

Our goal is to compute the gluon radiation field sourced by a set of interacting spinning particles satisfying the above equations of motion.   For our purposes in this paper, it is sufficient to compute the relevant observables to linear order in spins.   We solve the equations of motion as a perturbative expansion, formally\footnote{There are actually two different perturbative expansions for a system of particles with typical energy $E\gtrsim m$ and impact parameter $b$.   In the  limit relevant to the classical double copy $c^a\sim L=Eb\gg 1$, these two parameters coincide, with $\epsilon_{YM}\sim g_s^2 c^a \ll 1$ playing the role of the small expansion parameter.   See~\cite{Goldberger:2017vcg} for a more detailed discussion.}  in powers of the gauge coupling $g_s$, using the same method as in~\cite{Goldberger:2016iau}.    The starting point is the Yang-Mills equations, written in the gauge $\partial_\mu A^\mu_a=0$, 
\beq
\label{eq:gcurr}
\Box A^\mu_a=g_s{\tilde J}^\mu_a(x)=g_s J^\mu_a + g_s f^{abc} A^b_\nu(\partial^\nu A_c^\mu - F_c^{\mu\nu}),
\eeq   
where the current ${\tilde J}^\mu_a(x)$ is conserved, $\partial_\mu {\tilde J}^\mu_a(x)=0,$ but not gauge invariant.   Nevertheless, it is related to physical quantities measured by observers at infinity.   In particular, the long distance radiation field is related to the momentum space current ${\tilde J}^\mu_a(k) =\int d^d x e^{ik\cdot x} {\tilde J}^\mu_a(x)$, evaluated on-shell with $k^2=0$.   For example, in $d=4$ dimensions, the radiation field is given by
\beq
\label{eq:lda}
\lim_{r\rightarrow\infty}  A^a_\mu(x)  = {g_s\over 4\pi r}\int {d\omega\over 2\pi} e^{-i\omega t} {\tilde J}^\mu_a(k),
\eeq
with $k^\mu = (\omega,{\vec k})=\omega(1,{\vec x}/r)$, and similarly for general $d$.

\begin{figure}
\centering
\includegraphics[scale=0.3]{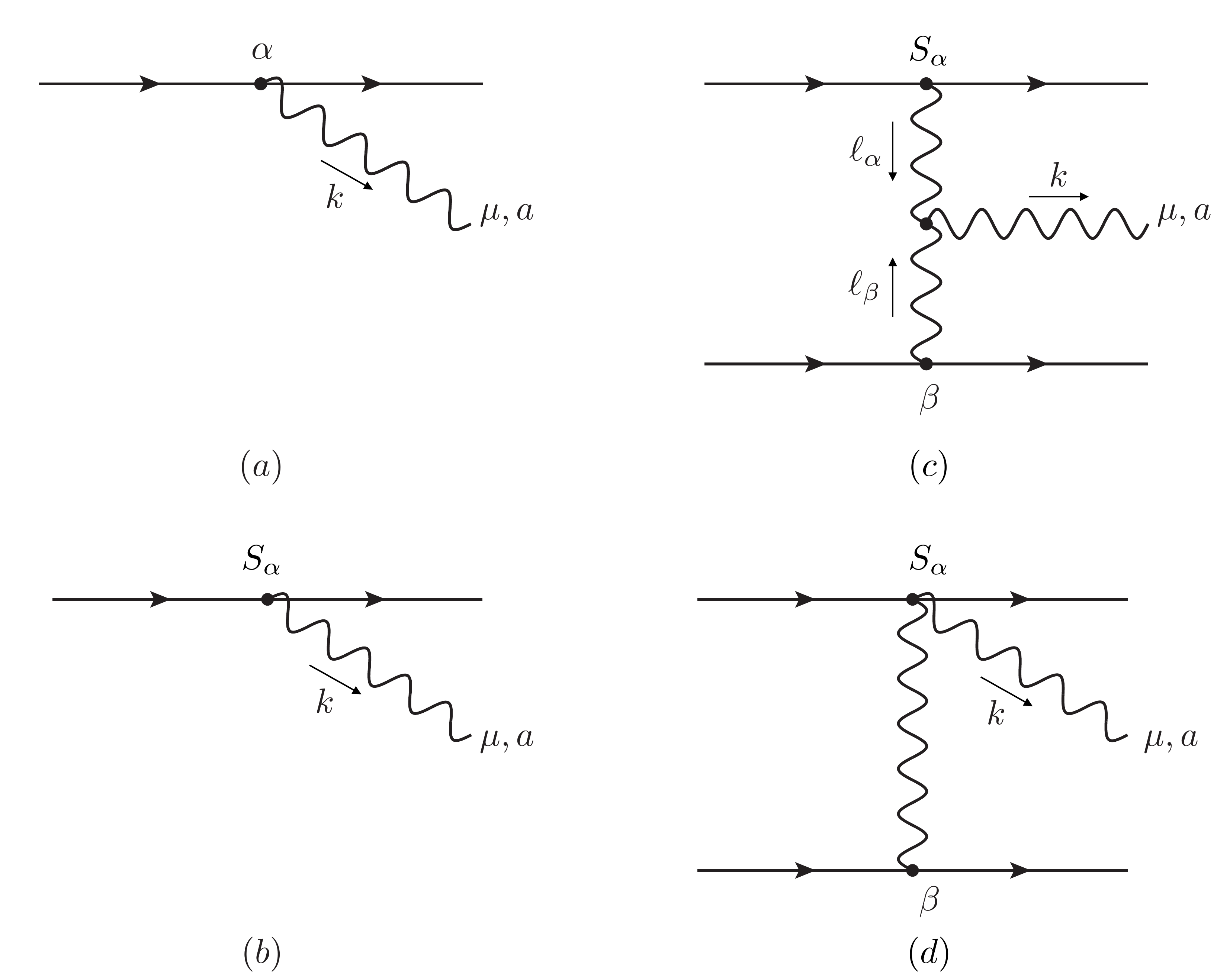}
\caption{Feynman diagrams for the perturbative expansion of ${\tilde J}^\mu_a(k)$ \label{fig:gluon1pt} up to order ${\cal O}(g^2_s)$.   The diagram $(a)$ represents corrections to the spin-independent color current due to the equations of motion.   Diagrams $(b)$-$(d)$ correspond to a single insertion of the spin-dependent color current.}    
\end{figure}

As long as the particles remain well separated, the current ${\tilde J}^\mu_a(k)$ can be calculated in perturbation theory, in terms of Feynman diagrams such as those (up to second order in the gauge coupling) shown in Fig.~\ref{fig:gluon1pt}.   These diagrams are computed using standard Yang-Mills Feynman rules, with insertions of the classical particle current Eq.~(\ref{eq:ppcurrent}).    The contribution from Figs.~\ref{fig:gluon1pt}(a) \& (b), to all orders in perturbation theory, can be written formally as
\begin{equation}
\label{eq:pc}
\left.J^\mu_a(k)\right|_{\mbox{Fig.}~\ref{fig:gluon1pt}(a)+(b)}= \sum_\alpha\int d\tau_\alpha\, e^{ik\cdot x_\alpha}c^a_\alpha \left[{v}^\mu_\alpha  +i \kappa_\alpha (S_\alpha\wedge k)^\mu\right].
\end{equation}
In this equation, we have abbreviated $x^\mu_\alpha=x^\mu_\alpha(\tau)$, $v^\mu_\alpha=v^\mu_\alpha(\tau)$, $c^a_\alpha = c^a_\alpha(\tau)$, $S^{\mu\nu}_\alpha=S^{\mu\nu}_\alpha(\tau)$ and introduced the notation  $(S_\alpha\wedge a)^\mu = S^{\mu\nu}_\alpha a_\nu$ for any Lorentz vector $a^\mu$.   To leading order in perturbation theory, the particles move on free trajectories with constant momentum $p^\mu_\alpha$ that is parallel to the velocity ${v}^\mu_\alpha$, implying that the spin $S^{\mu\nu}_\alpha$ is time-independent (see appendix ~\ref{sec:spapp}).    Thus we have at this order $x^\mu_\alpha = b^\mu_\alpha + p^\mu_\alpha \tau$, with constant $b^\mu_\alpha$, as well as ${\dot c}^a_\alpha=0$.   In this limit, the particles then source a static color current given by
\begin{eqnarray}
\label{eq:statcurr}
\left.{\tilde J}^\mu_a(k)\right|_{{\cal O}(g_s^0)} &=& \sum_\alpha  (2\pi)\delta(k\cdot p_\alpha)  e^{ik\cdot x_\alpha} c^a_\alpha \left[p^\mu_\alpha +i \kappa_\alpha (S_\alpha\wedge k)^\mu\right],
\end{eqnarray}
This static current cannot source radiation.   For on-shell gluons with $k^2=0$, $k\cdot p_\alpha$ is non-vanishing only if $p_\alpha$ is lightlike and collinear with $k$.    If $k$ is along the direction of $p_\alpha$, the second term in the above expression vanishes due to the constraint $(S_\alpha\wedge p_\alpha)^\mu=0$.   The first term also cannot contribute to the radiation amplitude ${\cal A}^a(k) = g_s \epsilon_\mu^*(k) {\tilde J}^\mu_a(k)$ since $p_\alpha^{\mu}$ dotted into the gluon polarization $\epsilon_\mu(k)$ is zero.   So, to get radiation, we must go to ${\cal O}(g_s^2)$.

At second order in perturbation theory, we need to account for two types of effects.  One is radiation emitted directly by the particles, depicted in Figs.~\ref{fig:gluon1pt}(a), (b).    The time-dependent current at this order in perturbation theory is conveniently computed by integrating by parts in Eq.~(\ref{eq:pc}) to put it in the form
\begin{equation}
J^\mu_a(k)= \sum_\alpha\int d\tau_\alpha e^{ik\cdot x_\alpha}{i\over k\cdot v_\alpha} \left[ {\dot c}^a_\alpha \left\{{v}^\mu_\alpha+i\kappa_\alpha ({\dot S}_\alpha\wedge k)\right\}+c^a_\alpha\left\{{\dot v}^\mu_\alpha +i\kappa_\alpha ({\dot S}_\alpha\wedge k)- {k\cdot {\dot v}_\alpha\over k\cdot v_\alpha} \left({v}^\mu_\alpha+i\kappa_\alpha (S_\alpha\wedge k)^\mu\right) \right\}\right].
\end{equation}
Here, the time evolution of the worldline degrees of freedom is due to their interaction with the field sourced by all the other particles,
\begin{equation}
\label{eq:lofield}
A^a_\mu(x) = g_s \int {d^d\ell\over (2\pi)^d} {e^{-i\ell\cdot x}\over\ell^2} {J}^\mu_a(\ell)= g_s \sum_\alpha\int d\tau_\alpha \, {d^d\ell\over (2\pi)^d} {e^{-i\ell\cdot (x- x_\alpha)}\over\ell^2} c^a_\alpha \left[{v}^\mu_\alpha +i \kappa_\alpha (S_\alpha\wedge\ell)^\mu \right].
\end{equation}
This is then fed into the equations of motion for the degrees of freedom $(p^\mu_\alpha, S^{\mu\nu}_\alpha,c_\alpha^a)$.    The spin-independent parts of the equations of motion were obtained in ref.~\cite{Goldberger:2016iau}, which we quote:
\begin{eqnarray}
{\dot p}^\mu_\alpha &=& ig_s^2 \sum_\beta \int d\tau_\beta  {d^d\ell\over (2\pi)^d} {e^{-i\ell\cdot x_{\alpha\beta}}\over\ell^2} (c_\alpha\cdot c_\beta) \left[(p_\alpha\cdot p_\beta) \ell^\mu - (\ell\cdot p_\alpha) p^\mu_\beta\right]\\
{\dot c}^a_\alpha &=& -ig_s^2 \sum_\beta \int d\tau_\beta  {d^d\ell\over (2\pi)^d} {e^{-i\ell\cdot x_{\alpha\beta}}\over\ell^2} [c_\alpha,c_\beta]^a (p_\alpha\cdot p_\beta),
\end{eqnarray}
where we have used the fact that in our worldline parametrization $p^\mu_\alpha={v}^\mu_\alpha$ up to terms that are higher order in perturbation theory.   At linear order in the spins, we have, from Eq.~(\ref{eq:ceom}),
\begin{eqnarray}
\left.{\dot c}^a_\alpha\right|_{{\cal O}(S^1)} 
&=& g_s^2 \sum_\beta \int d\tau_\beta {d^d\ell\over (2\pi)^d} {e^{-i\ell\cdot x_{\alpha\beta}}\over\ell^2} [c_\alpha,c_\beta]^a \left[\kappa_\alpha (\ell\wedge p_\beta)_\alpha-\kappa_\beta (\ell\wedge p_\alpha)_\beta\right],
\end{eqnarray}
with $(a\wedge b)_\alpha\equiv a\cdot (S_\alpha\wedge b)$.   Similarly, inserting the field Eq.~(\ref{eq:lofield}) into the Lorentz force law yields the result
\begin{eqnarray}
\label{eq:p}
\left.{\dot p}^\mu_\alpha\right|_{{\cal O}(S^1)} &=& -{g_s^2} \sum_\beta  \int d\tau_\beta  {d^d\ell\over (2\pi)^d} {e^{-i\ell \cdot x_{\alpha\beta}}\over\ell^2}  (c_\alpha\cdot c_\beta) \left[\kappa_\alpha  (\ell\wedge p_\beta)_\alpha \ell^\mu  - \kappa_\beta\left((\ell\wedge p_\alpha)_\beta \ell^\mu + (\ell\cdot p_\alpha) (S_\beta\wedge \ell)^\mu\right)\right]. 
\end{eqnarray} 
In the presence of a background gauge field, the momentum is no longer parallel to the velocity ${v}^\mu$.    The relation between these variables can be obtained by imposing that the constraint $p_\mu S^{\mu\nu}=0$ is consistent with the equations of motion, with the result
\begin{equation}
{v}^\mu = p^\mu + {g_s \over m_\alpha^2} (1+\kappa) c_a \, p^\sigma F^a_{\sigma\rho} S^{\rho\mu} + {\cal O}(S^2)
\end{equation}
to linear order in spin.    Then, Eq.~(\ref{eq:p}) implies that the spin correction to the velocity is
\begin{eqnarray}
\label{eq:x}
\left. {v}^\mu_\alpha\right|_{{\cal O}(S^1)} &=& p^\mu_\alpha + {ig_s^2(1+\kappa_\alpha)\over m_\alpha^2}\sum_\beta \int d\tau_\beta   {d^d\ell\over (2\pi)^d}   {e^{-i\ell \cdot x_{\alpha\beta}}\over\ell^2}  (c_\alpha\cdot c_\beta)\left[ \left((p_\alpha\cdot p_\beta) (S_\alpha\wedge\ell) - (\ell\cdot p_\alpha) (S_\alpha\wedge p_\beta)\right)^\mu\right].
\end{eqnarray} 
Notice that the orbital position is not well-defined in the massless limit $m_\alpha^2\rightarrow 0$ unless the chromomagnetic moment takes the special value $\kappa=-1$.  This value $\kappa=-1$ that ensures a smooth massless limit corresponds to the ``natural" magnitude, in the sense defined in ref.~\cite{Ferrara:1992yc}, of the gyromagnetic ratio $g$ of the particle.   In the particular case $d=4$, the non-relativistic limit of the the chromomagnetic coupling in Eq.~(\ref{eq:sint}) reduces to   
\begin{equation}
-{g_s\kappa\over m} \int dt \, c_a \, {\vec S}\cdot {\vec B}^a
\end{equation}
after accounting for the relation $m d\tau =ds$ between proper time and our worldline parametrization.   We see that $\kappa=-1$ corresponds to a massive classical particle, with spin $|{\vec S}|\gg \hbar$, whose gyromagnetic ratio corresponds to the Dirac value $g_D=2$.   We will also see $\kappa=-1$ playing an important role in the double copy in the next section.   Inserting Eqs.~(\ref{eq:p}),~(\ref{eq:x}) into the spin equation of motion gives, to linear order in spin, 
\begin{eqnarray}
\nonumber
\left. {\dot S}^{\mu\nu}_\alpha\right|_{{\cal O}(S^1)} &=&i{g^2_s} \sum_\beta \int d\tau_\beta   {d^d\ell\over (2\pi)^d}   {e^{-i\ell \cdot x_{\alpha\beta}}\over\ell^2}  (c_\alpha\cdot c_\beta) \left[\kappa_\alpha \left( \ell^\mu (S_\alpha\wedge p_\beta)^\nu- p_\beta^\mu (S_\alpha\wedge l)^\nu\right)\right.\\
& &{}  \left.+ {\left(1+\kappa_\alpha\right)\over m_\alpha^2}\left((p_\alpha\cdot p_\beta) (S_\alpha\wedge \ell) - (p_\alpha\cdot \ell) (S_\alpha\wedge p_\beta)\right)^\nu p_\alpha^\mu -(\mu\leftrightarrow \nu) \right].
\end{eqnarray}

%

The second type of contribution to the radiation field at infinity is due to the self-interactions of the gauge field.  These terms are conveniently organized in terms of the Feynman diagrams shown in Figs.~\ref{fig:gluon1pt}(c),(d).   We find
\begin{equation}
\left.{\tilde J}^\mu_a(k)\right|_{\mbox{Fig.}~\ref{fig:gluon1pt}(c)}=  i g_s^2 \sum_{\alpha,\beta} \kappa_\alpha [c_\alpha,c_\beta]^a \int d\mu_{\alpha\beta}(k) \ell_\alpha^2  (S_\alpha\wedge p_\beta)^\mu,
\end{equation}
\begin{equation}
\left.{\tilde J}^\mu_a(k)\right|_{\mbox{Fig.}~\ref{fig:gluon1pt}(d)} = i g_s^2 \sum_{\alpha,\beta} \kappa_\alpha [c_\alpha,c_\beta]^a \int d\mu_{\alpha\beta}(k) \left[ -2 (\ell_\alpha \wedge  \ell_\beta)_\alpha p_\beta^\mu  - 2 (k\cdot p_\beta) (S_\alpha\wedge \ell_\alpha)^\mu  + (\ell_\alpha\wedge p_\beta)_\alpha (\ell_\beta-\ell_\alpha)^\mu\right],
\end{equation}
where we have introduced the integration measure 
\beq
\label{eq:mudef}
d\mu_{\alpha\beta}(k) = d\tau_\alpha d\tau_\beta \left[{d^d\ell_\alpha\over (2\pi)^d} {e^{i\ell_\alpha\cdot x_\alpha}\over\ell^2_\alpha}\right] \left[{d^d\ell_\beta\over (2\pi)^d}{e^{i\ell_\beta\cdot x_\beta}\over\ell^2_\beta}\right] (2\pi)^d \delta^d(k-\ell_\alpha-\ell_\beta)
\eeq
over both worldline parameters and momenta.

We can now combine the effect of the time-dependent orbits with the contributions of the non-linear interactions in Fig.~\ref{fig:gluon1pt}(c),~(d)  to obtain the total current at ${\cal O}(g^2_s)$ and linear order in the spins.   The  result can be expressed as a sum of two color structures
\begin{eqnarray}
\left.{\tilde J}^\mu_a(k)\right|_{{\cal O}(S^1)} =i g_s^2  \sum_{\alpha,\beta}\int d\mu_{\alpha\beta}(k) \left[(c_\alpha\cdot c_\beta) c^a_\alpha {\cal A}^\mu_s+  [c_\alpha,c_\beta]^a {\cal A}_a^\mu\right],
\end{eqnarray}
with
\begin{eqnarray}
\label{eq:aadj}
\nonumber
{\cal A}^\mu_a &=& \kappa_\alpha \left[(\ell_\alpha\wedge p_\beta)_\alpha (\ell_\beta-\ell_\alpha)^\mu -  {\ell_\alpha^2\over k\cdot p_\alpha}  (\ell_\beta\wedge p_\beta)_\alpha p_\alpha^\mu -{\ell_\beta^2\over k\cdot p_\beta}  (\ell_\alpha\wedge p_\beta)_\alpha p_\beta^\mu + \ell_\alpha^2 (S_\alpha\wedge p_\beta)^\mu\right]\\
\nonumber\\
& & {}  -2 \kappa_\alpha(k\cdot p_\beta) \left[(S_\alpha\wedge \ell_\alpha)^\mu - {(k\wedge\ell_\alpha)_\alpha\over k\cdot p_\beta} p^\mu_\beta\right]
- \kappa_\alpha {\ell_\alpha^2\over k\cdot p_\alpha} (p_\alpha\cdot p_\beta) (S_\alpha\wedge k)^\mu.
\end{eqnarray}
and
\begin{eqnarray}
\label{eq:as}
\nonumber
{\cal A}_s^\mu &=&    {(1+\kappa_\alpha)^2 \over m_\alpha^2}\ell_\alpha^2  \left[  (k\cdot p_\alpha)\left\{(S_\alpha\wedge p_\beta)^\mu - {(k\wedge p_\beta)_\alpha\over k\cdot p_\alpha} p_\alpha^\mu\right\} + (p_\alpha\cdot p_\beta)\left\{(S_\alpha\wedge \ell_\beta)^\mu - {(k\wedge\ell_\beta)_\alpha\over k\cdot p_\alpha} p_\alpha^\mu\right\} \right]\\
\nonumber\\
\nonumber
& & -\kappa_\beta\ell_\alpha^2\left[(S_\beta\wedge\ell_\beta)^\mu- {(k\wedge\ell_\beta)_\beta\over k\cdot p_\alpha} p_\alpha^\mu \right]\\
\nonumber\\
\nonumber 
& & {} + \kappa_\alpha^2  { \ell_\alpha^2 \over k\cdot p_\alpha} \left[(k\cdot p_\beta)\left\{(S_\alpha\wedge \ell_\beta)^\mu - {(k\wedge\ell_\beta)_\alpha\over k\cdot p_\beta}  p_\beta^\mu \right\}  - (k\cdot\ell_\beta)\left\{ (S_\alpha\wedge p_\beta)^\mu - {(k\wedge p_\beta)_\alpha\over k\cdot p_\beta} \ell_\beta^\mu\right\}\right]\\
\nonumber\\
\nonumber
& &  {} + \kappa_\alpha {\ell_\alpha^2\over k\cdot p_\alpha}\left[(\ell_\beta\wedge p_\beta)_\alpha\left\{\ell_\beta^\mu - {k\cdot\ell_\beta\over k\cdot p_\alpha} p_\alpha^\mu\right\}   +(k\cdot p_\beta) (S_\alpha\wedge k)^\mu \right] +\kappa_\beta  {\ell_\alpha^2\over k\cdot p_\alpha} (\ell_\beta\wedge p_\alpha)_\beta\left[\ell_\beta^\mu - {k\cdot\ell_\beta\over k\cdot p_\alpha} p^\mu_\alpha\right]\\
& & {} -   \kappa_\alpha {\ell_\alpha^2\over (k\cdot p_\alpha)^2} (p_\alpha\cdot p_\beta) (k\cdot\ell_\beta) (S_\alpha\wedge k)^\mu
\end{eqnarray}
The analytic structure of the terms in this expression reflects their origin in Fig.~\ref{fig:gluon1pt}.   For example, the double pole at $\ell_\alpha^2=\ell_\beta^2=0$ is the contribution from the diagram in Fig.~\ref{fig:gluon1pt}(c), while the poles at $k\cdot p_\alpha=0$ generally correspond to the time dependence of the particles in orbital, color, and spin space.   It is straightforward to check that $k_\mu {\tilde J}_a^\mu(k)=0,$ so that we have obtained a consistent solution to the classical Yang-Mills equations for sources in general, but self-consistent, time-dependent orbits.

\section{Double copy}
\label{sec:dc}

We now apply the classical double copy rules proposed in~\cite{Goldberger:2016iau}, as applied to orbits with general time dependence in~\cite{Goldberger:2017vcg}.   In the spin-independent case, it was shown that the formal substitution rules
\begin{eqnarray}
\label{eq:mr}
\nonumber
c^a_\alpha(\tau) &\mapsto&  i p_\alpha^\mu(\tau),\\
f^{abc} c^a_\alpha c^b_\beta &\mapsto&  {1\over 2}\left[(p_\alpha\cdot p_\beta) (\ell_\beta-\ell_\alpha)^\nu + p_\beta\cdot (\ell_\alpha+q) p_\alpha^\nu-  p_\alpha\cdot (\ell_\beta+q) p_\beta^\nu \right],\\
\nonumber
p^\mu_\alpha(\tau) &\mapsto& p^\mu_\alpha(\tau),
\end{eqnarray}
together with  $g_s \mapsto {1/2 m^{(d-2)/2}_{Pl}}$, map the current ${\tilde J}^\mu_a(k)\mapsto i{\tilde T}^{\mu\nu}(k)$, to an object whose form is
 \begin{eqnarray}
 \label{eq:Tmunu}
\nonumber
{\tilde T}^{\mu\nu}(k) &=& {1\over 4 m_{Pl}^{d-2}} \sum_{\alpha,\beta} \int d\mu_{\alpha\beta}(k)\left[\left( {1\over 2} (p_\alpha\cdot p_\beta) (\ell_\beta-\ell_\alpha)^\nu + (k\cdot p_\beta) p_\alpha^\nu-  (k\cdot p_\alpha) p_\beta^\nu\right) {\cal A}^\mu_{adj}\right. \\
& &\hspace{5cm} {}\left. - (p_\alpha\cdot p_\beta) {p^\nu_\alpha} {{\cal A}}^\mu_{s}\right],
\end{eqnarray}
where
\begin{equation}
\left.{\cal A}^\mu_{adj}\right|_{{\cal O}(S^0)} =  (p_\alpha\cdot p_\beta)\left[{1\over 2} (\ell_\beta-\ell_\alpha)^\mu +{\ell_\alpha^2\over k\cdot p_\alpha} p_\alpha^\mu \right]+ (k\cdot p_\beta) p^\mu_\alpha - (k\cdot p_\alpha) p^\mu_\beta
\end{equation}
and 
\begin{equation}
\left. {{\cal A}}^\mu_s\right|_{{\cal O}(S^0)} = -{\ell_\alpha^2\over k\cdot p_\alpha} \left[(p_\alpha\cdot p_\beta) \left({1\over 2}(\ell_\beta-\ell_\alpha)^\mu -  {k\cdot\ell_\beta\over k\cdot p_\alpha} p^\mu_\alpha\right) -(k\cdot p_\alpha) p^\mu_\beta +(k\cdot p_\beta)p^\mu_\alpha\right].
\end{equation}
The effective source ${\tilde T}^{\mu\nu}(k)$ defined by Eq.~(\ref{eq:Tmunu}) is symmetric, ${\tilde T}^{\mu\nu}(k)={\tilde T}^{\nu\mu}(k)$, and for on-shell $k^2=0$ satisfies the Ward identity $k_\mu {\tilde T}^{\mu\nu}(k)=0$.  Therefore, it defines consistent graviton, ${\epsilon_{\mu\nu} (k)} {\tilde T}^{\mu\nu}(k)$, and scalar, ${\tilde T}^{\mu}{}_\mu(k)$, emission amplitudes\footnote{We choose normalization conventions in which the canonically normalized graviton emission amplitude is ${\cal A}_g = -{\epsilon_{\mu\nu}(k)} {\tilde T}^{\mu\nu}(k)\Big/\left(2 m_{Pl}^{{(d-2)}/ 2}\right)$.},  or equivalently radiation fields at retarded time $t$ and $r\rightarrow\infty$ (taking $d=4$ for illustration) 
\begin{eqnarray}
h_{\pm}(t,{\vec n}) &=&  {4G_N\over r} \int {d\omega\over 2\pi} e^{-i\omega t} \epsilon^*{}_{\pm}^{\mu\nu}(k) {\tilde T}_{\mu\nu}(k),\\
\phi(t,{\vec n}) &=&  {G_N\over r} \int {d\omega\over 2\pi} e^{-i\omega t} {\tilde T}^{\mu}{}_\mu(k),
\end{eqnarray}
in a theory of gravity coupled to point sources.  Here ${\vec n}={\vec k}/|{\vec k}|$ is the unit vector that points from the source to a far away detector, and $\omega=k^0$ is the frequency of radiation (in $d=4$, we define $G_N=1/32\pi m_{Pl}^2$).    

Ref.~\cite{Goldberger:2016iau}  verified by direct calculation, in the case of classical scattering and bremsstrahlung, that indeed ${\tilde T}^{\mu\nu}(k)$ matches the radiation fields in dilaton gravity coupled to point particles, with Lagrangian
\beq
\label{eq:gaction}
S_g= -2 m_{Pl}^{d-2}\int d^d x \sqrt{g} \left[R -(d-2) g^{\mu\nu}\partial_\mu\phi\partial_\nu\phi\right]-\sum_\alpha m_\alpha \int d\tau e^\phi.
\eeq
As shown in~\cite{Goldberger:2017vcg}, Eq.~(\ref{eq:Tmunu}) also captures the radiation fields of dilaton gravity for more general perturbative orbits, including non-relativistic bound systems.    

In the case of spinning particles, it is natural to apply the same color-to-kinematics replacements, as well as the mapping $S^{\mu\nu}_\alpha\mapsto S^{\mu\nu}_\alpha$.   This now yields an effective source ${\tilde T}^{\mu\nu}(k)$ which has the same form as in Eq.~(\ref{eq:Tmunu}), where at linear order in the spins ${\cal A}^\mu_s$ and ${\cal A}^\mu_a$ are given in Eqs.~(\ref{eq:aadj}),~(\ref{eq:as}).    In the spinning case,  ${\tilde T}^{\mu\nu}(k)$ is no longer symmetric.    Even though it satisfies $k_\mu {\tilde T}^{\mu\nu}(k)=0$, for generic values of the chromomagnetic parameters $\kappa_\alpha$, the spin-dependent source has $k_\nu {\tilde T}^{\mu\nu}(k)\neq 0$, so it does not yet define a consistent radiating solution in a theory of gravity coupled to other massless fields.   By simple inspection, $k_\nu {\tilde T}^{\mu\nu}(k)$ contains a term whose integrand is of the form 
\begin{equation}
{1\over 2} (p_\alpha\cdot p_\beta) (\ell_\alpha^2-\ell_\beta^2)  {(1+\kappa_\alpha)^2 \over m_\alpha^2}\ell_\alpha^2  \left[  (k\cdot p_\alpha)\left\{(S_\alpha\wedge p_\beta)^\mu - {(k\wedge p_\beta)_\alpha\over k\cdot p_\alpha} p_\alpha^\mu\right\} + (p_\alpha\cdot p_\beta)\left\{(S_\alpha\wedge \ell_\beta)^\mu - {(\ell_\alpha\wedge\ell_\beta)_\alpha\over k\cdot p_\alpha} p_\alpha^\mu\right\} \right],
\end{equation}
which, for generic $\kappa_\alpha,$ is not cancelled by other terms in $k_\nu {\tilde T}^{\mu\nu}(k)$.   Thus a necessary condition for consistency of the double copy is that the chromomagnetic coupling of each particle takes on the specific value\footnote{Ref.~\cite{Holstein:2006ry} noticed, in $d=4$, a similar factorization of graviton Compton scattering off massive particles with spin $s=1/2,1$.    For the special case in which the  particles have  $\kappa=-1$ or gyromagnetic ratio $g=2$, graviton scattering factorizes into two copies of photon scattering.}
\begin{equation}
\kappa_\alpha = -1.
\end{equation}
Once this choice is made, there are additional non-trivial cancellations among terms proportional to $\kappa_\alpha$ and $\kappa_\alpha^2$ that ensure  $k_\mu {\tilde T}^{\mu\nu}(k)=k_\nu {\tilde T}^{\mu\nu}(k)=0$ for $k^2=0$.  It follows that for the specific choice $\kappa_\alpha =-1$, the double copy rules define a consistent radiation field that includes, as before, graviton and scalar channels.  Because ${\tilde T}^{\mu\nu}(k) \neq {\tilde T}^{\nu\mu}(k)$, this theory also describes radiation into an antisymmetric mode, with amplitude  
\begin{equation}
{\cal A}_B = -{1\over 2 m_{Pl}^{(d-2)/2}} a^*_{\mu\nu}(k) {\tilde T}^{\mu\nu}(k),
\end{equation}
 where the (normalized) polarization tensor $a_{\mu\nu}(k) = - a_{\nu\mu}(k)$, $k^\mu a_{\mu\nu}(k)=0$,  is defined up to gauge transformations $a_{\mu\nu}(k)\rightarrow a_{\mu\nu}(k) + k_\mu \zeta_\nu(k) - k_\nu \zeta_\mu(k)$.   Turning on the particle spins, we can now probe the entire spectrum of the gravitational double copy, consisting of $h_{\mu\nu}$, $\phi$ as well as the Kalb-Ramond~\cite{Kalb:1974yc} axion (two-form gauge field) $B_{\mu\nu}(x)=-B_{\nu\mu}(x)$.   As in the non-spinning case, this theory is local.   In particular, the double pole in ${\tilde T}^{\mu\nu}(k)$ at $\ell_\alpha^2=\ell_\beta^2=0$, which encodes the tri-linear interactions of the fields in the gravitational sector, is analytic in momenta.
 
The form of the theory containing $(h_{\mu\nu},\phi,B_{\mu\nu})$ fields is largely fixed by general covariance and by the two-form gauge invariance $[\delta B(x)]_{\mu\nu} = [d\zeta(x)]_{\mu\nu}$, with one-form gauge parameter $\zeta_\mu$.   At the two-derivative level, the most general bulk Lagrangian is
\begin{equation}
\label{eq:sb}
S_g= -2 m_{Pl}^{d-2}\int d^d x \sqrt{g} \left[R -(d-2) g^{\mu\nu}\partial_\mu\phi\partial_\nu\phi + {1\over 12} f(\phi) H_{\mu\nu\sigma} H^{\mu\nu\sigma} \right],
\end{equation}
where $H_{\mu\nu\sigma} = (d B)_{\mu\nu\sigma}$ and $f(\phi)=1 + f'(0)\phi+\cdots $ is a function that, to linear order in $\phi$, we will fix below by comparing to the double copy prediction.    Because there is radiation in the axion channel, there must be a linear interaction with the point particle sources.    If the point particles carry no worldline degrees of freedom other than momentum and spin, the unique possibility at leading order in a derivative expansion is 
\begin{equation}
\label{eq:ppb}
S_{pp} = \int dx^\mu  {\tilde\kappa}(\phi) H_{\mu\nu\sigma} S^{\nu\sigma},
\end{equation}
 for some dilaton dependent function ${\tilde \kappa}(\phi)={\tilde\kappa} + {\tilde\kappa}'\phi+\cdots$, which we also determine to linear order in $\phi$.

 \begin{figure}
\centering
\includegraphics[scale=0.35]{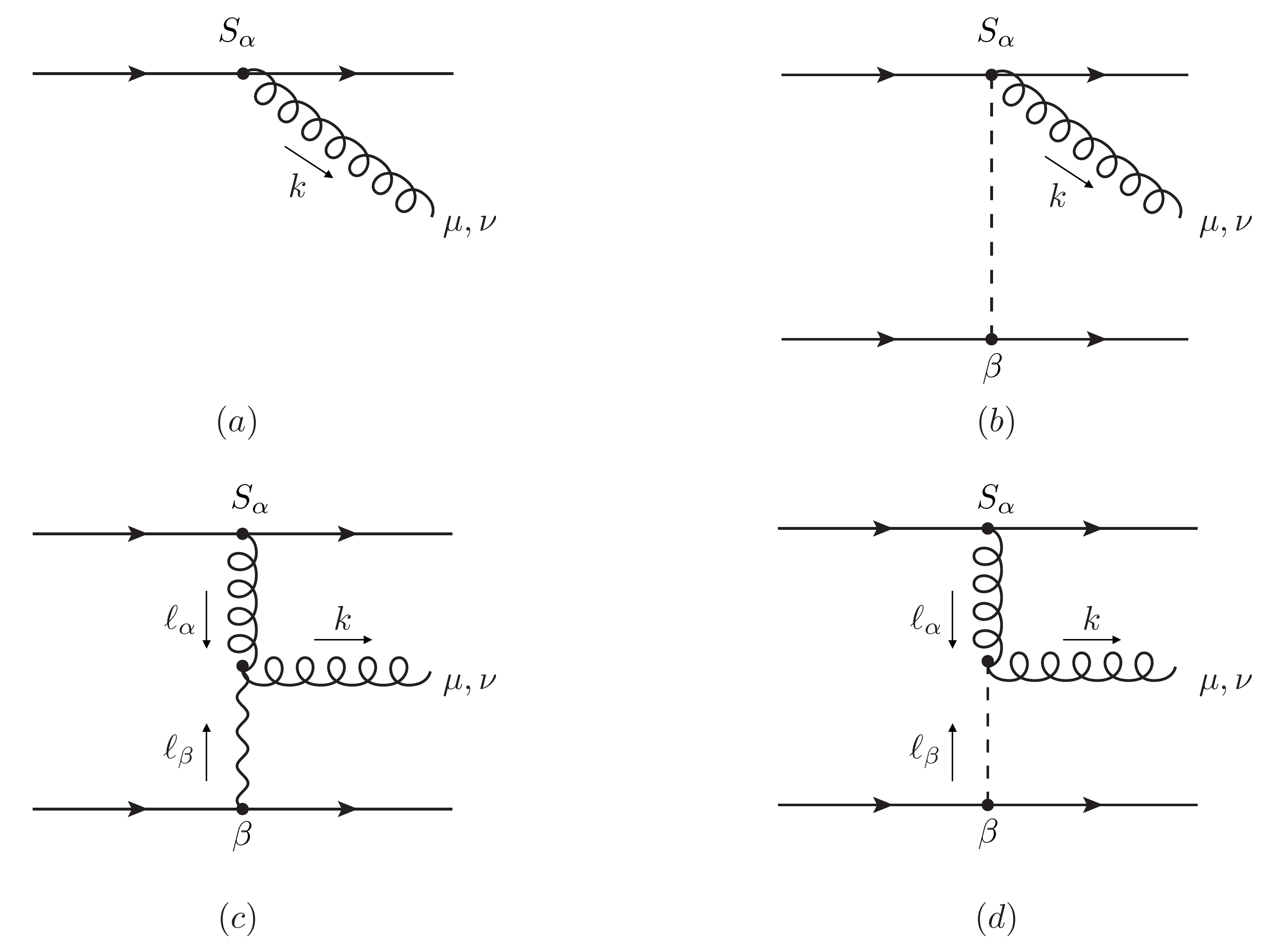}
\caption{Perturbative corrections to axion emission.  Dashed lines, wavy lines, and curly lines denote respectively scalar, graviton, and axion propagators\label{fig:axion1pt}.}    
\end{figure} 

To fix the form of the dilaton couplings in Eqs.~(\ref{eq:sb}),~(\ref{eq:ppb}), we directly compute the on-shell axion current ${\tilde J}^{\mu\nu}(x)=-{\tilde J}^{\nu\mu}(x),$ to linear order in both spin and the couplings ${\tilde\kappa}_\alpha$.   This current is defined such that the axion equations of motion take the form
\begin{equation}
\Box B^{\mu\nu}(x) = {1\over m_{Pl}^{d-2}} {\tilde J}^{\mu\nu}(x),
\end{equation}
which is a flat space wave equation, with $\Box=\partial_\mu \partial^\mu$ and index raising/lowering with the flat metric.   We write this equation in the gauge $\partial^\mu B_{\mu\nu}=0$, so that our current by definition obeys the Ward identity $\partial_\mu {\tilde J}^{\mu\nu}(x)=0$.   Given our normalization conventions, the amplitude for axion emission is then ${\cal A}_B = a^*_{\mu\nu}(k) {\tilde J}^{\mu\nu}/m_{Pl}^{(d-2)/2}$.

The expansion of the current $ {\tilde J}^{\mu\nu}(x)$ is depicted by the  graphs in Fig~\ref{fig:axion1pt}.   The diagram in Fig.~\ref{fig:axion1pt}(a), corresponding to direct emission from the particle worldlines, only contributes to radiation if the sources are time-dependent, and yields a contribution to the Kalb-Ramond current of the form
\begin{equation}
\left.{\tilde J}^{\mu\nu}(k)\right|_{\mbox{Fig.}~\ref{fig:axion1pt}(a)} = i\sum_\alpha \int ds {\tilde\kappa}_\alpha e^{i k\cdot x_\alpha} \left[ (k\cdot v_\alpha) S^{\mu\nu}_\alpha + v_\alpha^\mu (S_\alpha\wedge k)^\nu - v_\alpha^\nu (S_\alpha\wedge k)^\mu\right].
\end{equation}
To calculate this explicitly, we need to insert the solution to the ${\cal O}({\tilde\kappa}^0_\alpha)$ equations of motion for the orbital and spin degrees of freedom.   To linear order in spin, these are just the geodesic equation for $p^\mu = {\dot x}^\mu + {\cal O}(S^2)$ in the conformally scaled metric ${\tilde g}_{\mu\nu} = e^{2\phi} g_{\mu\nu}$, as well as the equation $v\cdot {\tilde\nabla} S^{\mu\nu}\approx 0$, where ${\tilde\nabla}_\lambda {\tilde g}_{\mu\nu}=0$.   Note that in Eq.~(\ref{eq:ppb}), we implicitly defined $S^{\mu\nu}$ to be the spin measured with respect to ${\tilde g}_{\mu\nu}$.   It is related to the spin in the Einstein frame by $S^{\mu\nu} = e^{-2\phi} S^{\mu\nu}_E$.   See appendix~\ref{sec:spapp} for details.    Inserting the leading order metric ${\tilde g}_{\mu\nu} = \eta_{\mu\nu} + {\tilde h}_{\mu\nu}$,
\begin{equation}
{\tilde h}_{\mu\nu}(x) = {1\over 2 m_{Pl}^{d-2}} \sum_\alpha\int d\tau_\alpha  {d^d\ell\over (2\pi)^d} {e^{-i\ell\cdot(x-x_\alpha)}\over\ell^2} p^\mu_\alpha p^\nu_\alpha,
\end{equation}
the equations of motion take the form
\begin{equation}
{\dot p}^\mu_\alpha = -{1\over 4 m_{Pl}^{d-2}}\sum_\beta \int d\tau_\beta  {d^d\ell\over (2\pi)^d} {e^{-i\ell\cdot(x_{\alpha\beta})}\over\ell^2} \left[(p_\alpha\cdot p_\beta)^2 \ell^\mu - 2 (p_\alpha\cdot p_\beta) (\ell\cdot p_\alpha) p^\mu_\beta\right]
\end{equation}
(from ref.~\cite{Goldberger:2016iau}), and 
\begin{equation}
{\dot S}^{\mu\nu}_{\alpha} = {i\over 4 m_{Pl}^{d-2}}\sum_\beta \int d\tau_\beta  {d^d\ell\over (2\pi)^d} {e^{-i\ell\cdot(x_{\alpha\beta})}\over\ell^2}\left[(p_\alpha\cdot\ell) p^\mu_\beta (S_\alpha\wedge p_\beta)^\nu -(p_\alpha\cdot p_\beta) p^\mu_\beta (S_\alpha\wedge\ell)^\nu - (p_\alpha\cdot p_\beta) \ell^\mu (S_\alpha\wedge p_\beta)^\nu - (\mu\leftrightarrow\nu)\right].
\end{equation}
Using these results, the direct emission term in {Fig.}~\ref{fig:axion1pt}(a) becomes
\begin{eqnarray}
\nonumber
\left.{\tilde J}^{\mu\nu}(k)\right|_{\mbox{Fig.}~\ref{fig:axion1pt}(a)}&=&{1\over 4 m_{Pl}^{d-2}} \sum_{\alpha\beta} {\tilde\kappa}_\alpha \int d\mu_{\alpha\beta}(k){\ell_\alpha^2}
\left[- {(p_\alpha\cdot p_\beta)\over k\cdot p_\alpha} ((k\cdot p_\alpha) p^\nu_\beta -(k\cdot p_\beta) p^\nu_\alpha)\left((S_\alpha\wedge \ell_\alpha)^\mu -{(k\wedge\ell_\alpha)_\alpha\over k\cdot p_\alpha} p^\mu_\alpha\right)\right. \\
\nonumber
& & {} + \left((k\cdot p_\alpha) p^\nu_\beta -(k\cdot p_\beta) p^\nu_\alpha + (p_\alpha\cdot p_\beta)\left(\ell_\alpha^\nu - {k\cdot \ell_\alpha\over k\cdot p_\alpha} p^\nu_\alpha\right)\right)\left(\left\{(S_\alpha\wedge p_\beta)^\mu - {(k\wedge p_\beta)_\alpha\over k\cdot p_\alpha} p^\mu_\alpha\right\}\right.       \\
& & {} \left. \left. + {(p_\alpha\cdot p_\beta)\over k\cdot p_\alpha} (S_\alpha\wedge k)^\mu\right) - (\mu\leftrightarrow\nu)\right].
\end{eqnarray}

The contribution of the diagram Fig.~\ref{fig:axion1pt}(b) is given by
\begin{eqnarray}
\label{eq:seag}
\left.{\tilde J}^{\mu\nu}(k)\right|_{\mbox{Fig.}~\ref{fig:axion1pt}(b)}=-{1\over 4m_{Pl}^{d-2}} \sum_{\alpha\beta}  \int d\mu_{\alpha\beta}(k) {m_\beta^2 {\tilde\kappa'}_\alpha\over d-2} \ell^2_\alpha\left[(k\cdot p_\alpha) S^{\mu\nu}_\alpha  + p^\mu_\alpha (S_\alpha\wedge k)^\nu -  p^\nu_\alpha (S_\alpha\wedge k)^\mu \right].
\end{eqnarray}
Finally, to compute Figs.~\ref{fig:axion1pt}(c),~(d), we need the  $B$-field propagator in the gauge $\partial^\mu B_{\mu\nu}=0$, 
\begin{equation}
\langle B_{\mu\nu}(k) B_{\rho\sigma}(-k)\rangle= {i\over 2 m_{Pl}^{(d-2)} k^2} \left[\eta_{\mu\rho} \eta_{\nu\sigma} - \eta_{\mu\sigma}\eta_{\nu\rho}\right].
\end{equation}
We then find that the relevant diagrams contributing to the axion current are 
 \begin{eqnarray}
 \label{eq:jeebs}
 \nonumber
\left.{\tilde J}^{\mu\nu}(k)\right|_{\mbox{Fig.}~\ref{fig:axion1pt}(c)+(d)}&=&{1\over 2 m_{Pl}^{d-2}} \sum_{\alpha\beta} {\tilde\kappa}_\alpha \int d\mu_{\alpha \beta}(k) \left[{m_\beta^2  (4+f'(0)) \over 2(d-2)} (k\cdot p_\alpha)\left(\ell^\nu_\alpha-{k\cdot\ell_\alpha\over k\cdot p_\alpha} p^\nu_\alpha\right) \left((S_\alpha\wedge\ell_\alpha)^\mu - {(k\wedge\ell_\alpha)_\alpha\over k\cdot p_\alpha} p^\mu_\alpha\right)\right.\\
\nonumber
& &{}+(k\cdot p_\beta) \left((k\cdot p_\beta) p^\nu_\alpha - (k\cdot p_\alpha) p^\nu_\beta -(p_\alpha\cdot p_\beta) \left(\ell_\alpha^\nu - {k\cdot \ell_\alpha\over k\cdot p_\beta} p_\beta^\nu\right)\right) \left((S_\alpha\wedge\ell_\alpha)^\mu -{(k\wedge\ell_\alpha)_\alpha\over k\cdot p_\beta} p^\mu_\beta\right)\\
& & {} \left.+(\ell_\alpha\wedge p_\beta)_\alpha \left((k\cdot p_\beta) p^\nu_\alpha - (k\cdot p_\alpha) p^\nu_\beta\right)\left(\ell^\mu_\alpha -{k\cdot\ell_\alpha\over k\cdot p_\alpha} p_\alpha^\mu\right) -(\mu\leftrightarrow\nu)\right].
 \end{eqnarray}
 
 We can now compare the gravity result with the double copy prediction.   We see from Eqs.~(\ref{eq:seag}),~(\ref{eq:jeebs}) that cancelling the explicit dependence on the dimension $d$ of the gravity Feynman rules requires the choice of parameters
 \begin{eqnarray}
f'(0) = -4, & \tilde{\kappa}_\alpha'=0.
 \end{eqnarray}
 We also find, by brute force calculation, cancellations only for the special case in which the particles have universal axion couplings 
\begin{equation}
{\tilde \kappa}_\alpha ={1\over 4}.
\end{equation}

For this choice of parameters, the difference between the anti-symmetric double copy amplitude,  ${\cal A}_B=-a_{\mu\nu}(k) {\tilde T}^{\mu\nu}(k)\Big/\left(2 m_{Pl}^{(d-2)/2}\right)$ and the axion emission amplitude computed directly using Eq.~(\ref{eq:sb}) takes the form
 \begin{eqnarray}
\nonumber
 a_{\mu\nu}(k)\left[{\tilde J}^{\mu\nu}(k) - {1\over 2}{\tilde T}^{[\mu\nu]}(k)\right] &=& {a_{\mu\nu}(k)\over 4 m_{Pl}^{d-2}}\sum_{\alpha\beta} \int d\mu_{\alpha\beta}(k)\left[
(p_\alpha\cdot p_\beta) \left\{\ell^2_\beta p^\nu_\beta (S_\alpha\wedge\ell_\alpha)^\mu - \ell^2_\alpha p^\nu_\alpha (S_\beta\wedge\ell_\beta)^\mu)-(\mu\leftrightarrow\nu)\right\}\right.\\
\nonumber
& & {}\hspace{1cm}\left. -  (\ell^2_\beta (\ell_\alpha\wedge p_\beta)_\alpha + \ell^2_\alpha (\ell_\beta\wedge p_\alpha)_\beta) \left(p^\mu_\alpha p^\nu_\beta-p^\nu_\alpha p^\mu_\beta\right)\right],\\
 \end{eqnarray}
after dropping contributions that vanish when $k^2=0$, or when dotted into the on-shell polarization tensor $a_{\mu\nu}(k)$.   The integrand is anti-symmetric under label exchange $\alpha\leftrightarrow\beta$, while the measure $\sum_{\alpha\beta}\int d\mu_{\alpha\beta}(k)$ is symmetric.   We therefore find precise agreement between the double copy and the axion emission amplitude computed directly in the gravity theory.

Note that the agreement between the two results is only operative for specific parameter choices.    On the gauge theory side, the chromomagnetic coupling must take on the value ${\kappa}_\alpha=-1$.   Otherwise, the double copy amplitude obtained by applying the rules in Eq.~(\ref{eq:mr}) is not consistent with gravitational Ward identities.     Once this choice is made, the gravity theory must have couplings among $\phi$, $B_{\mu\nu}$ and the graviton that ensure the cancellation of any explicit dependence on spacetime dimensionality $d$ introduced by the graviton propagator.   The couplings we found are consistent with the bulk gravity action in Eq.~(\ref{eq:st}), and point-particle interactions 
 \begin{equation}
 \label{eq:app}
S_{pp} = {1\over 4} \int dx^\mu  H_{\mu\nu\sigma} S^{\nu\sigma}.
 \end{equation}
written in terms of the ``string frame'' spin.    In particular, Eq.~(\ref{eq:st}) is equivalent, after Weyl re-scaling to the string frame, to the action for the massless bosonic gravitational sector of non-critical string theory (ignoring the gauge field and ${\cal O}(\hbar)$ contributions to the dilaton potential).    As discussed in~\cite{Bern:1999ji} (see also ref.~\cite{Scherk:1974mc}), the form of these couplings ensures the cancellation of explicit $d$-dependence in scattering amplitudes computed using Eq.~(\ref{eq:st}).    Of course, to check the consistency of the linear spin-dependent terms in the double copy, it is also important to verify that radiation in the graviton and scalar channels agrees with the prediction of Eq.~(\ref{eq:st}).    This is indeed the case, and the details of those results will be presented elsewhere~\cite{JS}.

\section{Discussion and conclusions \label{sec:conclusion}}

In this paper, we have extended the perturbative double copy to the case of spinning particles.    The same color-to-kinematics replacement rules introduced in~\cite{Goldberger:2016iau}, and generalized in~\cite{Goldberger:2017vcg}, map the long-distance field produced by a collection of spinning color charges to a corresponding classical solution in the gravity theory of coupled scalar, axion, and graviton modes.  One novel aspect of the calculation is that, unlike the case of spinless particles, the double copy map only yields a consistent solution that satisfies gravitational Ward identities if the sources on the gauge theory side have specific values of the chromomagnetic dipole interaction.  In $d=4$, this corresponds to the ``natural''~\cite{Ferrara:1992yc} Dirac value $g=2$ of the gyromagnetic ratio.    Once this choice is made, the theory on the gravity side is consistent with the action of string gravity.   This action, which also arises as the BCJ double copy of pure Yang-Mills, has the form~\cite{Scherk:1974mc} of an Einstein-Hilbert type action $S_g=-2 m_{Pl}^{d-2}\int d^d x \sqrt{{\tilde g}} R({\tilde g},{\hat\nabla})$ with a non-Riemannian connection whose torsion is related to the axion field strength $H^\mu{}_{\rho\sigma}$.

The double copy also fixes the strength of the spin-axion interaction on the gravity side.   It would be interesting to see if these point-particle interactions arise as limits of more fundamental classical extended objects on either side of the correspondence\footnote{On the gauge theory side, a possible candidate UV completion of the spinning color charge with $g=2$ is a classical open bosonic string with endpoint Chan-Paton charges coupled to  $A^\mu_a$.   Taking the point particle limit of this object then yields a worldline action for the center-of-mass coordinate of the string which includes the chromomagnetic term with coefficient $\kappa=-1$, in agreement with results in~\cite{Ferrara:1992yc}.   However, the obvious double copy of this object, namely an oriented closed string with worldsheet interaction  $\int d^2\sigma\epsilon^{ab} B_{\mu\nu}(X)\partial_a X^\mu \partial_b X^\nu$ does not have a point particle limit consistent with Eq.~(\ref{eq:app}).   Rather, the closed string's axion dipole interaction is of the form ${1\over 4} \int dx^\mu H_{\mu\rho\sigma}\left(S_L^{\rho\sigma}-S_R^{\rho\sigma}\right)$, where the relative sign between left-moving and right-moving (string frame) spins is due to worldsheet parity $L\leftrightarrow R$.   See~\cite{Garcia-Saenz:2017wzf} for a systematic discussion of the multipole expansion for the Kalb-Ramond current.  The fact that the axion does not couple to the total spin $S^{\mu\nu}=S_L^{\mu\nu}+S_R^{\mu\nu}$ reflects the lack of rotational invariance of the closed string configuration, so that, unlike a $d=4$ Kerr black hole, its multipole moments are not fully determined by the spin.}.    On the gravity side, the extended objects in question are presumably spinning black holes, or perhaps naked singularities~\cite{Luna:2016hge}, with non-zero dilaton monopole and axion dipole charges.   While vacuum solutions to Eq.~(\ref{eq:st}) with these exact properties are not known, ref.~\cite{Campbell:1990ai} constructed spinning, axion-dilaton solutions of $d=4$ massless string theory, with axion and dilaton ``hair" sourced by ${\cal O}(\alpha')$ string corrections to the action in Eq.~(\ref{eq:st}).

Further constraints on the structure of the objects that arise from the double copy would result by including, for instance, higher order spin corrections to the results presented here.   In particular, to get a consistent double copy at ${\cal O}(S^2)$, it may be necessary to include worldline couplings to gravity of the schematic form $\int d\tau R_{\mu\nu\rho\sigma} S^{\mu\nu} S^{\rho\sigma}$, which encodes the quadrupole moment induced by the rotation (with definite coefficients~\cite{Porto:2008jj} for a $d=4$ Kerr black hole).   Also, applying the double copy to classical scattering solutions with gluon radiation in the initial state would test possible worldline terms of the schematic form $\int d\tau R_{\mu\nu\rho\sigma}^2$, $\int d\tau H_{\mu\nu\rho}^2$ that encapsulate the ``tidal'' responses of the extended object in the point particle limit.     Finally, the spin results here can also be extended in the direction of the classical double copy~\cite{KS2,BDC,Goldberger:2017frp} between Yang-Mills solutions and the bi-adjoint scalar theory of~\cite{Cachazo:2013iea}.   We hope to address some of these questions in future work.

\section{Acknowledgments}

WG thanks A. Nicolis for discussions on Kalb-Ramond multipoles, and P. Daamgard and L. Stein for respectively bringing refs.~\cite{Ferrara:1992yc,Campbell:1990ai} to his attention. This research was partially supported by Department of Energy grant DE-FG02-92ER-40704.  

\begin{appendix}

\section{Formalism for classical spinning particles}
\label{sec:spapp}
In this appendix, we provide a self-contained review of the spin formalism used in this paper.   The approach we follow is equivalent to the one introduced in~\cite{Porto:2005ac,Porto:2006bt,Porto:2008tb,Porto:2008jj} in the context of worldline effective theories of gravity~\cite{Goldberger:2004jt}, which itself is based on the classic papers~\cite{Papapetrou:1951pa,Dixon:1970zza,Hanson:1974qy}.

\subsection{The free spinning particle}
 We begin with free particles moving in flat space, and define the system in terms of a worldline $x^\mu(s)$ (with $s$ an arbitrary parameter) and an einbein $e(s)$ to enforce worldline reparametrization invariance, $s\rightarrow s'(s)$, $e'(s') ds' = e(s) ds$.   In addition we introduce~\cite{Hanson:1974qy} an orthonormal reference frame $e^I_{\mu}(s)$ moving along with the particle, as well as its inverse $e^{\mu}_I(s)$.   These are related by the constraints
\begin{eqnarray}
\label{eq:orthoc}
\eta_{IJ} e^I_{\mu} e^J_{\nu} = \eta_{\mu\nu} & \eta_{\mu\nu} e^{\mu}_I e^{\nu}_J = \eta_{IJ}.
\end{eqnarray}
The rotation of the the particle relative to fixed inertial frames is then encoded in the angular velocity 
\begin{equation}
\Omega^{IJ} = \eta^{\mu\nu} e^I_\mu {d\over ds}    e^J_\nu = -\Omega^{JI}.
\end{equation}
Finally, we also introduce worldline degrees of freedom $p_I(s)$ and $S^{IJ}(s)$ corresponding to momentum and spin (which can be regarded as conjugate variables to $x^\mu(s)$ and $\Omega^{IJ}$. 

The system of variables $(x^\mu,e,e^I_\mu,p_I,S^{IJ})$ is redundant, and constraints must be imposed to reduce the number of independent degrees of freedom down to the physical number implied by Poincare invariance.   A common choice in the literature (which we follow) is to impose the constraint
\begin{equation}
\label{eq:ssc}
S^{IJ} p_J =0,
\end{equation}
sometimes referred to as the ``covariant spin supplementary condition.''   As discussed in ref~\cite{Porto:2005ac,Porto:2006bt}, the variational principle is essentially fixed by worldline reparametrization and Lorentz invariance to be of the form $\delta S_{pp}=0$, where
\begin{equation}
\label{eq:Spp}
S_{pp}= -\int dx^\mu e^I_\mu p_I +{1\over 2} \int ds S^{IJ} \Omega_{IJ} + {1\over 2}\int ds e\left(p_I p^I - m^2(S) + \cdots\right) + \int ds e \lambda_I S^{IJ} p_J.
\end{equation}
Here we have introduced a Lagrange multiplier $\lambda_I$ to enforce the constraint in Eq.~(\ref{eq:ssc}).    The equations of motion follow from the variation of $S_{pp}$ with respect to $(x^\mu,e,e^I_\mu,p_I,S^{IJ},\lambda_I)$.    Varying with respect to $x^\mu$ yields
\begin{equation}
{d\over ds} p^\mu = {d\over ds}(e^\mu_I p^I)=0,
\end{equation}
while varying with respect to $p_I$ implies
\begin{equation}
\label{eq:dp}
e^{-1}{\dot x}^\mu= p^\mu - \lambda_\nu S^{\mu\nu},
\end{equation}
with ${\dot x}^\mu = dx^\mu/ds$.

The variation with respect to $e^I_\mu$ must be performed in a way that is consistent with the orthonormality constraint Eq.~(\ref{eq:orthoc}), so we introduce
\begin{equation}
\theta^{IJ} = \eta^{\mu\nu}e^I_\mu \delta e^J_\nu = -\theta^{JI}.
\end{equation}
The vanishing of the coefficient of $\theta^{IJ}$ in the variation of $S_{pp}$ then implies
\begin{equation}
{d\over ds} S^{IJ} = \Omega^I{}_K S^{KJ} - \Omega^{J}{}_{K} S^{KI} + e^J_\mu {\dot x}^\mu p^I -  e^I_\mu {\dot x}^\mu p^J.
\end{equation}
This equation simplifies if expressed in terms of the spin measured in the inertial frame, $S^{\mu\nu} = e^\mu_I e^\nu_J S^{IJ},$ which becomes
\begin{equation}
{d\over ds}S^{\mu\nu} = {\dot x}^\nu p^\mu -  {\dot x}^\mu p^\nu.
\end{equation}

Variation with respect to $S^{IJ}$ provides the relation between spin and angular velocity
\begin{equation}
e^{-1}\Omega^{IJ} = \lambda^J p^I - \lambda^I p^J + {\partial\over \partial S_{IJ}} m^2(S).
\end{equation}
In general this is a model-dependent relation which is sensitive to the specific choice of `Regge trajectory' $m^2(S)$.   By definition, a spherically symmetric particle is one in which $m^2(S)$ is a function of the Lorentz invariants\footnote{In generic dimension $d$, a basis of such invariants consist of a finite set of traces of the anti-symmetric matrix $S^{IJ}$ .   More generally, if the particle is not spherically symmetric, $m^2$ can also depend on additional structure, fore instance the inertia tensor $I^{IJ}$ defined in the frame of the particle.   See~\cite{Delacretaz:2014oxa} for a coset space formulation of the action for extended objects without rotational symmetry.} constructed from $S^{IJ}$.   The function $m^2(S)$ is not predicted by the effective point particle theory, but is determined by matching to the full UV theory of the extended object.    We assume  $m^2(S)$ is an analytic function about $S^{IJ}=0$, so that the action can be expanded in powers of the polynomial invariants of $S^{IJ}$.   Note that if we treat $x^\mu,p^\mu$ and $S^{\mu\nu}$ as the fundamental variables, the equations of motion can be formulated without specific knowledge of this UV function $m^2(S)$.

Finally, variation with respect to the Lagrange multipliers $e, \lambda_I$ reproduces the constraints $S^{IJ} p_J=0$ and $p_I p^I = m^2(S)$.   From these constraints, we find that $p_\mu p^\mu = m^2(S)$ and $S_{\mu\nu} S^{\mu\nu}$ are constants of the motion.    Demanding consistency of the time evolution with the constraint $S^{\mu\nu} p_\nu=0$ gives
\begin{equation}
0= {d\over ds} (S^{\mu\nu} p_\nu) = p^2 {\dot x}^\mu - ({\dot x}\cdot p) p^\mu.
\end{equation}
Thus in the absence of external fields, $p^\mu$ is collinear with the vector ${\dot x}^\mu$ tangent to the worldline.  In light of Eq.~(\ref{eq:dp}), ${\dot x}\cdot p = e p^2$ and we find
\begin{equation}
p^\mu = e^{-1} {\dot x}^\mu.
\end{equation}
Thus, ${\dot S}^{\mu\nu}=0$, and the free particle moves in a straight line with constant momentum $p^\mu$ and constant spin $S^{\mu\nu}$.

\subsection{Coupling to pure gravity}

To include gravity, we covariantize the action Eq.~(\ref{eq:Spp}) to allow for general spacetime diffeomorphisms.   First, replace $\Omega^{IJ}$ with
\begin{equation}
\Omega^{IJ}\rightarrow \Omega^{IJ}=g^{\mu\nu} e^I_\mu v^\sigma \nabla_\sigma e^J_\nu  = g^{\mu\nu} e^I_\mu({\dot e}^J_\nu -v^\sigma\Gamma^\lambda{}_{\sigma\nu} e^J_\lambda),
\end{equation}
and promote the constraints on $e^I_\mu$ to 
\begin{eqnarray}
\label{eq:orthocg}
\eta_{IJ} e^I_{\mu} e^J_{\nu} = g_{\mu\nu}(x(s)) & g_{\mu\nu}(x(s)) e^{\mu}_I e^{\nu}_J = \eta_{IJ}.
\end{eqnarray}

Even in problems without gravity, the advantage of turning on a background metric is that it allows one to define a symmetric energy-momentum tensor
\begin{equation}
T^{\mu\nu}(x) = -{2\over \sqrt{g}} {\delta\over \delta g_{\mu\nu}(x)}S_{pp},
\end{equation}
which in the presence of gravity is conserved $\nabla_\nu T^{\mu\nu}(x)=0$ as a consequence of the Einstein equations.  It is well known~\cite{Papapetrou:1951pa,Dixon:1970zza} that the dynamics of the spinning point particle follow model-independently from the conservation of the distributional energy-momentum tensor defined by $S_{pp}$.     

To obtain the spinning particle energy-momentum tensor we vary $S_{pp}$, taking into account the dependence on $g_{\mu\nu}$ in Eq.~(\ref{eq:orthocg}),
\begin{equation}
\delta e^I_{\mu} = {1\over 2} e^I_\alpha g^{\alpha\beta} \delta g_{\beta\mu},
\end{equation}
as well as the variation of the connection 
\begin{equation}
g_{\sigma\lambda}\delta \Gamma^\lambda_{\mu\nu} = {1\over 2} (\nabla_\mu \delta g_{\sigma\nu} + \nabla_\nu \delta g_{\sigma\mu} - \nabla_\sigma \delta g_{\mu\nu}).
\end{equation}
From these formulas, we get  $S_{IJ}\delta \Omega^{IJ} = -S^{\mu\nu} v^\sigma \nabla_{[\mu} \delta g_{\nu]\sigma}$, and the spinning particle energy-momentum tensor is
\begin{equation}
T_{pp}^{\mu\nu}(x) = \int dx^{(\mu} p^{\nu)} {\delta(x-x(s))\over \sqrt{g}} + \int  dx^{(\mu}S^{\nu)\alpha}  \nabla_\alpha {\delta(x-x(s))\over \sqrt{g}},
\end{equation}
The meaning of the covariant derivative on the Dirac delta function is more fully explained in refs.~\cite{Dixon:1970zza,Bailey:1975fe,Mino:1995fm}.    Here, we only need to know that it can be integrated by parts covariantly against an arbitrary test vector field $X^\mu(x)$,
\begin{equation}
\int d^d x\sqrt{g} X^\lambda(x) \nabla_\lambda \left( {\delta(x-z)\over \sqrt{g}} \right)= - \nabla_\lambda X^\lambda(z).
\end{equation}
As explained in refs.~\cite{Dixon:1970zza,Bailey:1975fe,Mino:1995fm} the form of $T^{\mu\nu}$ gives physical meaning to the variables $p_\mu = e^I_\mu p_I$ and $S^{\mu\nu}$ obeying the constraint $S^{\mu\nu} p_\nu=0$.    For instance, turning off gravity and defining the total momentum and angular momentum measured in a fixed frame by
\begin{eqnarray}
P^\mu &=& \int d^3 {\bf x} T^{0\mu}({\bf x},x^0),\\
J^{\mu\nu} &=& \int d^3 {\bf x} x^{[\mu} T^{0\nu]}({\bf x},x^0),
\end{eqnarray}
yields $P^\mu=p^\mu$ and $J^{\mu\nu} = x^\mu p^\nu - x^\nu p^\mu + S^{\mu\nu}$.   More generally, in a curved background, $P^\mu$ and $J^{\mu\nu}$ can be defined as integrals over an observer dependent spacelike hypersurface, see~\cite{Mino:1995fm}.   These relations give the interpretation of $x^\mu$ as the `center-of-mass' worldline and $S^{\mu\nu}$, with $S^{\mu\nu} p_\nu =0$, as the particle spin relative to the center of mass.

We now obtain the spinning particle equations of motion by averaging $\nabla_\mu T_{pp}^{\mu\nu}=0$ against the test vector $X_\mu(x)$:
\begin{equation}
\int d^dx \sqrt{g} X_\mu \nabla_\nu T_{pp}^{\mu\nu} = -\int dx^{(\mu} p^{\nu)} \nabla_\nu X_\mu(x(s)) + \int dx^{(\mu} S^{\nu)\rho} \nabla_{\rho} \nabla_{\nu} X_\mu(x(s))
\end{equation}
Shuffling terms around using $[\nabla_\mu,\nabla_\nu]X_\rho = -X^\sigma R_{\sigma\rho\mu\nu}$ and the Bianchi identities for the Riemann tensor then gives
\begin{eqnarray}
\nonumber
\int d^dx \sqrt{g} X_\mu \nabla_\nu T_{pp}^{\mu\nu} &=&  \int ds X_\mu\left({\dot x}\cdot \nabla p^\mu +{1\over 2} {\dot x}^\sigma S^{\lambda\rho} R^\mu{}_{\sigma\lambda\rho}\right) - \int ds \nabla_{\nu} X_{\mu} \left({\dot x}^{[\mu} p^{\nu]}+{1\over 2}{\dot x}\cdot \nabla S^{\mu\nu}\right).\\
\end{eqnarray}
Since the test vector field $X^\mu(x)$ is arbitrary, the conservation of $T_{pp}^{\mu\nu}$ implies that the coefficients of $X_\mu(s)$ and $\nabla_\nu X_\mu(s)$ vanish independently in the above expression,
\begin{eqnarray}
({\dot x}\cdot \nabla) p^\mu &=& -{1\over 2} {\dot x}^\sigma S^{\lambda\rho} R^\mu{}_{\sigma\lambda\rho},\\
({\dot x}\cdot \nabla)S^{\mu\nu} &=& p^\mu {\dot x}^\nu - p^\nu {\dot x}^\mu, 
\end{eqnarray}
which are the Papapetrou-Mathison-Dixon equations of motion.   These equations together with the constraint $S^{\mu\nu} p_\nu=0$ determine the dynamics.    It follows in particular that $S_{\mu\nu} S^{\mu\nu}$ and $m^2=p_\mu p^\mu$ are constants of the motion.  Note that in general ${\dot x}^\mu$ and $p^\mu$ are not collinear.  Rather,
\begin{equation}
 e p^\mu = {\dot x}^\mu +{1\over 2 m^2} {\dot x}^\sigma R_{\nu\sigma\lambda\rho} S^{\mu\nu} S^{\lambda\rho}
\end{equation}
after using the constraints (including ${\dot x}\cdot p = e p^2$ from Eq.~(\ref{eq:dp})) and the above equations of motion.

\subsection{Dilaton gravity}

When the dilaton $\phi$ is included, the point particle action Eq.~(\ref{eq:Spp}), is modified in such a way that every term picks up an arbitrary coefficient function of $\phi$.   However, it is possible to perform dilaton-dependent redefinitions of the worldline variables $(e^I_\mu,p_I,S^{IJ},\Omega_{IJ},\lambda_I)$ such that the most general action takes the same form as in Eq.~(\ref{eq:Spp})
\begin{equation}
S_{pp}= -\int dx^\mu e^I_\mu p_I +{1\over 2} \int ds S^{IJ} \Omega_{IJ} + {1\over 2}\int ds e\left(p_I p^I - m^2(S,\phi) + \cdots\right) + \int ds e \lambda_I S^{IJ} p_J,
\end{equation}
where now the Regge function $m^2(S,\phi)$ becomes dilaton-dependent, and the $e^I_\mu$ now satisfy the constraint $\eta_{IJ} e^I_\mu e^J_\nu = {\tilde g}_{\mu\nu} = e^{2\phi} g_{\mu\nu}$.   It follows immediately that the equations of motion for $p^\mu$ and $S^{\mu\nu}$ are again of the form
\begin{eqnarray}
({\dot x}\cdot {\tilde\nabla}) p^\mu &=& -{1\over 2} {\dot x}^\sigma S^{\lambda\rho} {\tilde R}^\mu{}_{\sigma\lambda\rho},\\
({\dot x}\cdot {\tilde\nabla}) S^{\mu\nu} &=& p^\mu {\dot x}^\nu - p^\nu {\dot x}^\mu, 
\end{eqnarray}
where now ${\tilde\nabla}_\mu$ and ${\tilde R}^\mu{}_{\sigma\lambda\rho}$ are the covariant derivative and Riemann tensor corresponding to the conformal metric ${\tilde g}_{\mu\nu}$.    The relation between $p^\mu$ and the velocity is now
\begin{equation}
 e p^\mu = {\dot x}^\mu +{1\over 2 m^2} {\dot x}^\sigma {\tilde R}_{\nu\sigma\lambda\rho} S^{\mu\nu} S^{\lambda\rho}.
\end{equation}

\subsection{Gauge interactions}

The interactions of the spin with a dynamical gauge field are accomplished by modifying the action in Eq.~(\ref{eq:Spp}) to
\begin{equation}
S_{pp}\rightarrow S_{pp} - g_s\int dx^\mu  c_a A^a_\mu +{g_s\kappa\over 2} \int ds e  c_a S^{\mu\nu} F^a_{\mu\nu}+\cdots,
\end{equation}
where $c_a(s)$ is a color degree of freedom (``color charge") carried by the particle, $g_s$ is the gauge coupling constant, and $\kappa$ controls the strength of the chromo-magnetic dipole interaction.   As in the gravitational case, the dynamics follows from conservation laws
\begin{eqnarray}
D_\mu J^{\mu}_a &=& 0,\\
\nabla_{\mu} \left(T^{\mu\nu}_{YM} + T^{\mu\nu}_{pp}\right) &=& 0.
\end{eqnarray}
Here, $D_\mu =\nabla_\mu + i g_s T^a A_\mu^a$ is the gauge covariant derivative.  The color current is given by 
\begin{equation}
J_a^\mu(x) = -{1\over g_s} {\delta\over \delta A^a_\mu(x)} S_{pp} = \int dx^\mu c_a  {\delta(x-x(s))\over \sqrt{g}} -  {\kappa} \int ds e S^{\mu\alpha} D_{\alpha}\left(c_a  {\delta(x-x(s))\over \sqrt{g}}\right),
\end{equation}
and the Yang-Mills equations of motion are $D_\nu F^{\nu\mu}_a=g_s J^\mu_a$.   By considering the integral $0=\int d^d x \sqrt{g} X^a D_\mu J^\mu_a$ for $X^a(x)$ an arbitrary Lorentz scalar in the adjoint representation, we obtain
\begin{equation}
e^{-1}({\dot x}\cdot D) c^a = -{1\over 2} g_s\kappa S^{\mu\nu} f^{abc} F^b_{\mu\nu} c^c = {i \over 2} g_s\kappa [S^{\mu\nu} F_{\mu\nu},c]^a.
\end{equation} 
Unlike the case of spinless particles, the color charge is no longer parallel transported along the particle.   However, it is still true that the gauge invariant $c_a c^a$ is a constant of the motion.

The orbital equations of motion are obtained by including the coupling of the gauge field to the spin in $T^{\mu\nu}_{pp}$.   Using $\delta S^{\mu\nu} = {1\over 2} S^\mu{}_\lambda \delta g^{\lambda\nu}-{1\over 2} S^\nu{}_\lambda \delta g^{\lambda\mu}$, the correction to $T^{\mu\nu}_{pp}$ is 
\begin{equation}
T^{\mu\nu}_{pp}\rightarrow T^{\mu\nu}_{pp}-{\kappa g_s} \int ds e {\delta(x-x(s))\over \sqrt{g}} c_a {F^a}_\alpha{}^{(\mu}{} S^{\nu)\alpha}.
\end{equation}
From the Yang-Mills equations of motion  $\nabla_\nu T^{\mu\nu}_{YM}=-g_sF^{a\mu\nu}  J^a_\nu ,$ and thus by the same method as in gravity
\begin{eqnarray}
\nonumber
\int d^d x \sqrt{g} X_\mu \left(\nabla_\nu T^{\mu\nu}_{pp} - g_s F^{a\mu\nu} J^a_\nu \right) &=&  \int ds X_\mu\left({\dot x}\cdot \nabla p^\mu +{1\over 2} {\dot x}^\sigma S^{\lambda\rho} R^\mu{}_{\sigma\lambda\rho}-g_s c^a F^{\mu\nu}_a {\dot x}_\nu \right.\\
\nonumber
& & \left. +{1\over 2}\kappa g_s e c_a S^{\alpha\beta} D^\mu F^a_{\alpha\beta}\right) \\
& & - \int ds \nabla_{\nu} X_{\mu} \left({\dot x}^{[\mu} p^{\nu]}+{1\over 2}{\dot x}\cdot \nabla S^{\mu\nu} +g_s\kappa e c^a F^{\alpha[\mu}_a S_\alpha{}^{\nu]}\right),
\end{eqnarray}
after using the Bianchi identity, $D_\mu F_{\alpha\beta}+\mbox{cyclic}=0$.  A non-trivial check is that the coefficient of $\nabla_\nu X_\mu$ is automatically anti-symmetric, as a result of a cancellation between terms in $T^{\mu\nu}_{pp}$ and $J^\mu_a$.   Turning off gravity for simplicity, we arrive at
\begin{eqnarray}
{d\over ds} p^\mu &=&   g_s c^a F_a^{\mu\nu} {\dot x}_\nu -{e\over 2}\kappa g_s  c_a S^{\alpha\beta} D^\mu F^a_{\alpha\beta},\\
{d\over ds} S^{\mu\nu} &=&  p^\mu {\dot x}^\nu - p^\nu {\dot x}^\mu - 2\kappa g_s e c^a F^{\alpha[\mu}_a S_\alpha{}^{\nu]}.
\end{eqnarray}
It is straightforward to check that these equations also follow via variation of the action with respect to $(x^\mu,e^I_\mu,e,p_I,S^{IJ},\lambda_I)$ directly in flat space.   Using the same type of argument as in ref.~\cite{Bailey:1975fe}, one can show that these equations imply that $S_{\mu\nu} S^{\mu\nu}$ is conserved along the worldline.   In addition we have the conservation law
\begin{equation}
{d\over ds} \left(p_\mu p^\mu + g_s\kappa c_a S^{\mu\nu}F^a_{\mu\nu}\right) = 0,
\end{equation}
so that $m^2 = p_\mu p^\mu + g_s\kappa c_a S^{\mu\nu}F^a_{\mu\nu}$ plays the role of a (field-dependent) invariant mass.

\end{appendix}

\end{document}